\documentstyle[preprint,aps]{revtex}

\tightenlines
\begin{document}
\draft
\preprint{HEP/123-qed}
\title{Self Organization of Edge and Internal Pedestals in a Sandpile}
\author{S. C. Chapman$^1$\footnote{sandrac@astro.warwick.ac.uk},
}
\address{
$^1$Physics Dept. Univ. of Warwick,
Coventry CV4 7AL, UK
}
\author{
R. O. Dendy$^2$}
\address{ 
$^2$EURATOM/UKAEA Fusion Association, Culham Science Centre, Abingdon,
Oxfordshire OX14 3DB, United Kingdom}
\author{B. Hnat$^1$}
\date{\today}
\maketitle
\begin{abstract}

The temperature profiles of magnetically confined plasmas can display
distinctive longlived pedestals at the edge and internally. Here we
show that such structures can arise naturally through avalanching
transport in a sandpile model. A single control parameter that is constant
across the sandpile determines the occurrence and regularity of these
effects, as well as the entrainment of global confinement to edge
pedestal dynamics. The system dynamics indicate that the pedestals
are a consequence of an inverse cascade in real space, and that
self organization is necessary for their occurrence.
\end{abstract}

%\narrowtext
%\twocolumn

A distinctive feature of magnetically confined plasmas is that
they can sustain local regions having very steep (indeed, almost discontinuous)
temperature gradients. Pedestals in the edge temperature are a key
feature of the good confinement regimes of tokamaks (``H-modes");
for a recent review see Ref.\cite{hugill2000}.
Additionally, advanced operating regimes for tokamaks have now been
accessed, which involve the creation of ``internal transport barriers"
(ITBs) -- steplike features in the temperature profiles internal to the plasma,
see for example Refs.\cite{Synakowski98}-\cite{Kinsey01}.
The occurence of such structures in externally heated plasma systems that
are diffuse,
high temperature,
and turbulent is surprising, and is a striking instance of their capacity
for macroscopic self organization. Here we shall identify how similar
structures arise spontaneously in the sandpile model of
Ref.\cite{Chap:Row:2000}.
Because of the simplicity of this model, it is possible to formally
characterize and
explain the mechanisms underlying pedestal formation, and to identify
links to tokamak plasma behavior. This approach is complementary
to studies that employ largescale numerical simulation of tokamak plasmas,
which have had some success in illuminating the subtle interplay between
turbulence and bulk flows that may give rise to ITBs\cite{Kinsey01,Rogers00}.

The essential ingredients of the sandpile algorithm of
Ref.\cite{Chap:Row:2000} are: (i)randomized critical gradients which must be
exceeded locally before any transport can take place; (ii)central fuelling,
so that sand can only leave the system by means of systemwide avalanches; and
(iii)its only distinctive feature, namely a characteristic lengthscale
$L_f$ for
fast redistribution which may be considered a proxy for turbulent transport, as
discussed in Ref.\cite{prl}. Evidence for
avalanche-type transport from tokamak experiments\cite{rhodes99,Politzer2000}
and numerical simulations\cite{Garbet:Waltz,Sarazin:Ghendrih}
provides growing support for the
applicability in some circumstances of the sandpile paradigm introduced into
fusion plasma physics in recent
years\cite{Newman:Carreras96}-\cite{Hicks:Carreras:2001}.
Particularly relevant to the present study
are observations and analysis\cite{Carreras:vanM:98,Pedrosa:Hidalgo99} of
edge plasma turbulence in a range of
magnetically confined plasmas. These suggest that edge plasma turbulence -- the
environment in which H-mode edge pedestals form, which then coexist with
the turbulence -- self organizes into a critical state,
independent of the size and plasma characteristics of the
devices considered. It is known\cite{prl} that, depending on the value of
the control
parameter $L_f$, the statistical behavior of the sandpile model of
Ref.\cite{Chap:Row:2000}
displays features reminiscent of enhanced confinement phenomenology
in tokamaks. These include the time averaged height profiles, which possess
edge pedestals in the good confinement regime; furthermore
the frequency of systemwide avalanches resulting in mass loss
scales with stored sandpile energy in the same way as the frequency of edge
localized modes (ELMs) scales with stored energy in tokamaks, see Fig.6 of
Ref.\cite{prl}. We emphasize that $L_f$ is kept constant across the sandpile
in any given computational run, so that the critical gradient and
redistribution rules are identical at every cell of our sandpile, apart
from the initial and final cells. It follows that where pedestals arise,
they represent a true emergent phenomenon. In this respect our approach differs
essentially from that of Refs.\cite{Newman:Carreras96,Hicks:Carreras:2001},
where a local transport barrier is specified at the outset, by declaring the
critical gradient and redistribution rules to be different for a
specific local group of cells.

The three dimensional plot of Fig.1 shows sandpile height as a function of
position as time evolves, for the good confinement regime\cite{prl} with
$L_f=50$ in a 512 cell system. It displays two distinct phases.
First, there is
a relaxation phase where the sandpile profile is smooth down to the
self organized edge pedestal,
except within a distance $L_f$ of the core where fueling has a continual
local effect.
During the relaxation phase mass loss occurs via many systemwide avalanches
closely spaced in time, which carry sand over the sandpile boundary.
The relaxation phase terminates with a final systemwide avalanche, after which
the growth phase begins.
The growth phase is characterized by a stationary edge pedestal which resides at the
outermost cell
of the sandpile. As time progresses, additional pedestals (localized
regions of steep gradient
just below critical) form successively at positions increasingly close to
the core of the sandpile,
with average separation $\sim L_f$. Each of these is generated at
positions where (outward propagating) major
internal avalanches have come to rest. The location of the most recently formed
(and therefore innermost) internal
pedestal propagates inward during the growth phase.

Figure 2 is motivated by simultaneous multichannel measurements of tokamak
temperature
profiles in the presence of ITBs; see for example Fig.3 of
Ref.\cite{Burrell98} and Fig.1 of Ref.\cite{Kinsey01}, and also the results
of  numerical simulations shown in Fig.2 of Ref.\cite{Kinsey01}.
Figure 2 shows sandpile height at different
positions from the edge to the centre. Just over two growth and relaxation
phases are shown.
The successive formation of internal pedestals is reflected in a stepwise
increase in height at any given point during the growth phases.
Using this diagnostic, evidence for stepwise increments is clearest in the
region of the sandpile that lies between the core and the edge. Since the
internal
pedestals form at locations increasingly close to the core as the growth
phase proceeds, the points within the sandpile that are most affected by the
formation of these pedestals are those that are closer to the core of the
sandpile. Points nearer the edge are only affected by the formation of the
first few
internal pedestals during the early growth phase.
The results shown in Fig.2, which bear some resemblance to the
experimental and numerical results of  Refs.\cite{Burrell98,Kinsey01},
emerge naturally from the dynamics of the sandpile during the growth phase
of its good confinement regime. Central to this structure is the unexpected
capacity of this sandpile to
organize persistent steep pedestals both at the edge and internally.

The role of these internal pedestals and their relationship to the edge pedestal
is highlighted in Fig.3: here all cells at which the gradient exceeds $z_c/2$
are marked by black points, while all other cells are left blank.
Figure 3(a)
($L_f=50$) follows five of the growth and relaxation cycles
shown in Fig.2.
The edge pedestal is visible close to the sandpile boundary in both the
relaxation and the
growth phases. Its time behaviour is essentially regular and, as we shall
see, orders the
structure internal to the sandpile.
For any $L_f <N/4$, the location of each internal pedestal is fixed during a
given growth phase, so that they persist as distinguishable features of the
time averaged
phenomenology of the sandpile. In contrast, Fig.3(b) shows the behavior for the
poor confinement regime with $L_f=250 >N/4$.
This regime corresponds to a self organized critical (SOC) state,
see Ref.\cite{prl} and below.
Although the sandpile sucessively fills and empties, it does
so in an irregular manner. Pedestals can be seen both at the edge and
internal to the
sandpile, but these are no longer organized in a coherent pattern.

A resilient edge pedestal arises for all $L_f$; the pedestal is
steep, indeed unresolved, in that the entire change in height occurs between
neighboring cells. The location of the edge pedestal is
strongly time dependent for large $L_f$, whereas for small $L_f$
it is confined to a region close to the outermost cell of the sandpile.
The time averaged profile in the edge region
therefore depends strongly on $L_f$, and the steadiest edge pedestal
corresponds to
small $L_f$ and good confinement. Following each avalanche, in our algorithm
the value of the critical gradient $z_c$ is randomised about a mean value
at all cells that participated in the avalanche.
Nevertheless the gradient at the edge pedestal
remains always close to, but just below,
the critical value $z_c$ as shown in
Fig.4. Elsewhere in the sandpile (for example at its midpoint, see Fig.4),
except where internal pedestals arise, the gradient is well below $z_c$.
The internal pedestals appear as barriers to transport:
despite their apparent fragility against avalanching (gradient $z$ close to
$z_c$),
no sand passes through either the edge or the internal pedestals until the
final
avalanche that terminates the growth phase.

The physical mechanisms and principles underlying the self organization of
the edge pedestal and multiple internal pedestals that arise
in the sandpile model
of Refs.\cite{Chap:Row:2000,prl} are therefore of considerable interest.
As a first step, we confirm the hypothesis of Ref.\cite{prl} that the good
confinement regime (small $L_f$) corresponds to low dimensional behavior.
In Fig.5 the position of the last occupied cell
at time $t$ is plotted against that at time $t+\tau$, where $\tau =50$, for
runs with  $L_f = (a)50$, $(b)150$ and $(c)250$ in a 512 cell
system. This is an example of phase space reconstruction,
achieved here by embedding\cite{ott}.
Figure 5(a) shows low dimensional system dynamics that repeatedly
follow a simple limit cycle (attractor) around a restricted region of the
reconstructed phase space. This implies that the large number of cells in
the sandpile have
self organized: their collective dynamics are encapsulated by a small number of
dynamical variables. Once $L_f$ is increased to $150$ (Fig.5(b)), the
simple limit cycle seen in Fig.5(a) bifurcates, and more stochastic
behavior is seen in Fig.5(c) ($L_f=250$). Thus increasing complexity of the
phase
space portrait correlates with deterioration of confinement\cite{prl}.
The lowest confinement regime corresponds to selfsimilar
avalanche statistics. This is associated with a nontrivial fixed point
in the space of the parameter used \cite{tam} to perform
rescaling under the renormalization group procedure,
corresponding to behavior that is both self organized
and critical\cite{nature}. Importantly, global relaxation of the sandpile
is ultimately
achieved by large (systemwide) avalanches for all values of $L_f$.
When $L_f$ is of order the system size $N$, systemwide avalanches are
straightforwardly
propagated: because $L_f \sim N$, no characteristic scale
is imposed by the redistribution process and the dynamics are selfsimilar
and in SOC\cite{tam}.
Conversely, when $L_f$ is significantly distinct from the system size
(found empirically to be $L_f <N/4$),
there is scale breaking. This leads to broken power law avalanche
statistics\cite{Chap:Row:2000},
and the system is no longer in SOC. By separating the characteristic
lengthscales we also
effectively separate the longterm growth-relaxation timescale from the time
interval between systemwide avalanches. In the SOC regime, such a
distinction is not
possible. Thus, in the
good confinement regime (when $L_f <N/4$) the requirement for self
organization is satisfied.
The feature evolving on the slow timescale, namely the position of the edge
pedestal, is sufficient to determine the details of the internal dynamics.
It organizes the sequence of events leading to the
sucessive formation of internal edge pedestals and the time variation of
total energy (sand) in the system: in short, the sandpile is entrained to
its edge. The characteristic signature of the onset of
self organization is low dimensional dynamics,
seen in Figs.4($L_f=50$ traces) and 5(a), as opposed to the
irregular time evolution and selfsimilar statistical properties of
the system when self organised and critical
(Figs.4($L_f=250$ traces) and 5(c)).

The confinement physics of our sandpile model offers
a robust framework in which a distinctive  structure of edge
and internal pedestals (previously known only from tokamak plasmas)
arises naturally. There is only one control parameter,
$L_f/N$, which
can be considered as a proxy for the lengthscale of turbulent transport,
normalized to system size. Provided that this lengthscale
is sufficiently short, the underlying inverse cascade in real
space gives rise to persistent, marginally
subcritical profile steps whose formation point propagates inwards in
the growth phase. Furthermore the self
organized edge pedestal is continuously present in all phases of the
sandpile evolution, and positions itself
exactly at the sandpile boundary throughout the growth phase.
These results are sufficient to indicate that some of the distinctive edge
and internal pedestal phenomenology seen in tokamak plasmas can also arise
in a simpler
idealized confinement
system, and that they may be linked to the observed avalanching transport
phenomena. This suggests that quantitative studies to elucidate the balance
between
diffusive transport and avalanching transport in tokamaks would be fruitful.
The results further suggest a test of the depth of the physical analogy
that we have found, as follows. If the analogy is deep, there will exist
one or a few
dimensionless control parameters, linked to the properties of the turbulent
transport,
that entirely determine the key features
of the confinement phenomenology -- regular or irregular -- of tokamak
plasmas that we have cited here. For example, these parameters would
control the extent to which
global confinement is entrained to edge pedestal dynamics. The search for
such parameters, for
example by further application of the techniques of nonlinear time series
analysis to edge
plasma measurements as initiated in
Refs.\cite{Carreras:vanM:98,Pedrosa:Hidalgo99}, is potentially
highly rewarding.

\acknowledgments
We are grateful to George Rowlands and Jack Connor for comments and
suggestions. SCC was supported by a PPARC lecturer fellowship, ROD by
Euratom and the UK
DTI, and BH by HEFCE.

Captions

FIG 1: Three dimensional view of the sandpile height profile for $L_f=50$. Time
evolves along the partially hidden axis. One relaxation phase and  one
subsequent
growth phase are shown.

FIG 2: Local height of sand as a function of time at different
locations in the sandpile relative to the centre cell $n=1$: (a) n=20,
(b) n=65, (c) n=100, (d) n=125, (e) n=150 and (f) n=300.
System size $N=512$, control parameter $L_f=50$.

FIG 3: Location of cells where the value of the local gradient exceeds $z_c/2$
for (a)$L_f=50$ and (b)$L_f=250$. Zero corresponds to the apex where fueling
occurs. The edge pedestal is visible as the uppermost trace.

FIG 4: Local gradient normalized to the local value of
$z_c$ for cells at the edge ($\Delta_e$, upper traces, value
close to unity) and halfway
into the sandpile ($\Delta_{mid}$, lower traces, value close to zero),
for $L_f=$ (top) 250, (centre) 150, (bottom)
 50.

FIG 5: Phase space reconstruction of the dynamics of the edge position
$ep(t)$. Plotted are coordinates $ep(t)$ versus $ep(t+\tau)$ for $\tau=50$
and $L_f = (a)50$, $(b)150$ and $(c)250$. The system dynamics explore
a larger region of the phase space with increasing values of $L_f$.
The topology shown is insensitive to the value of $\tau$ in the range of
interest.

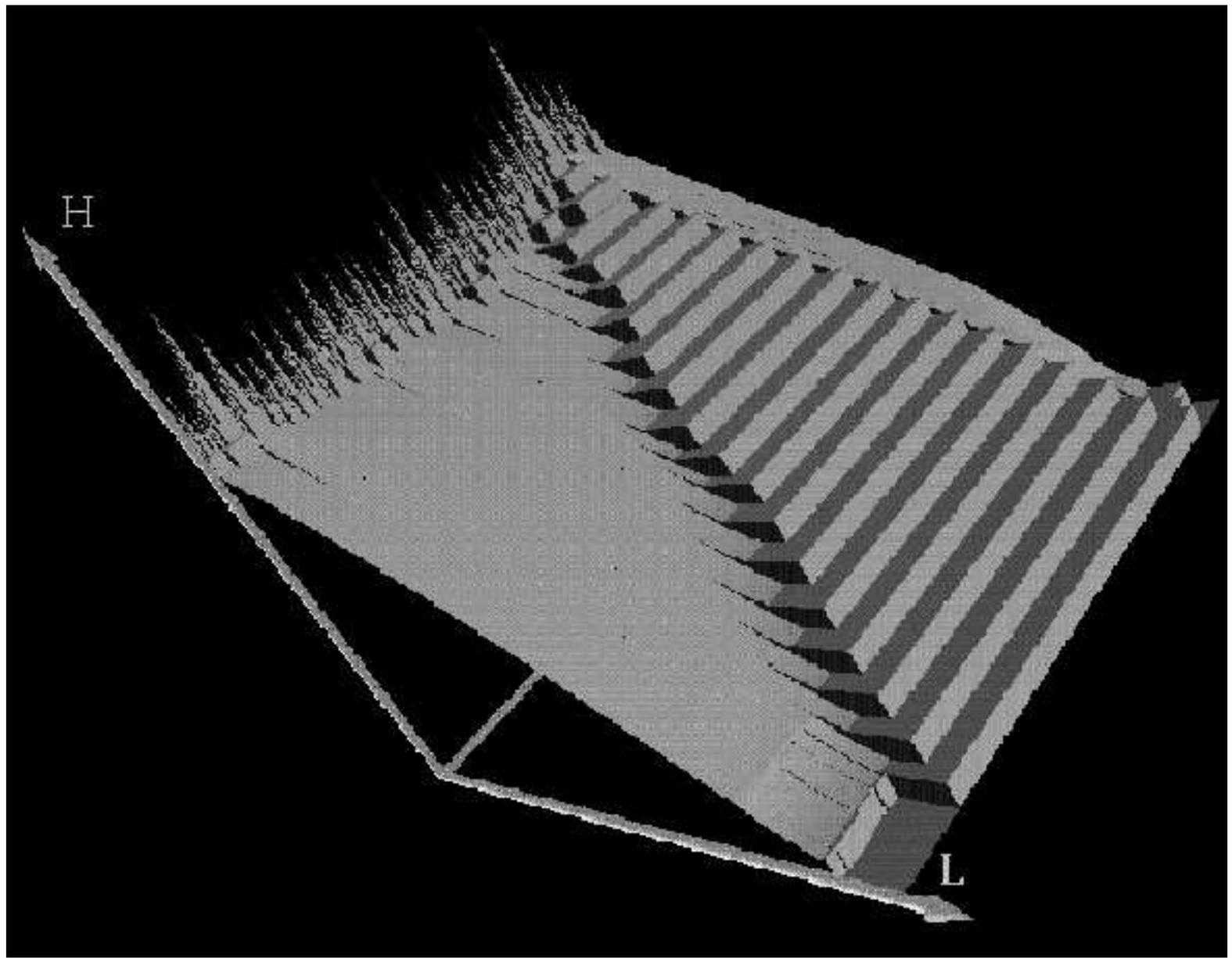
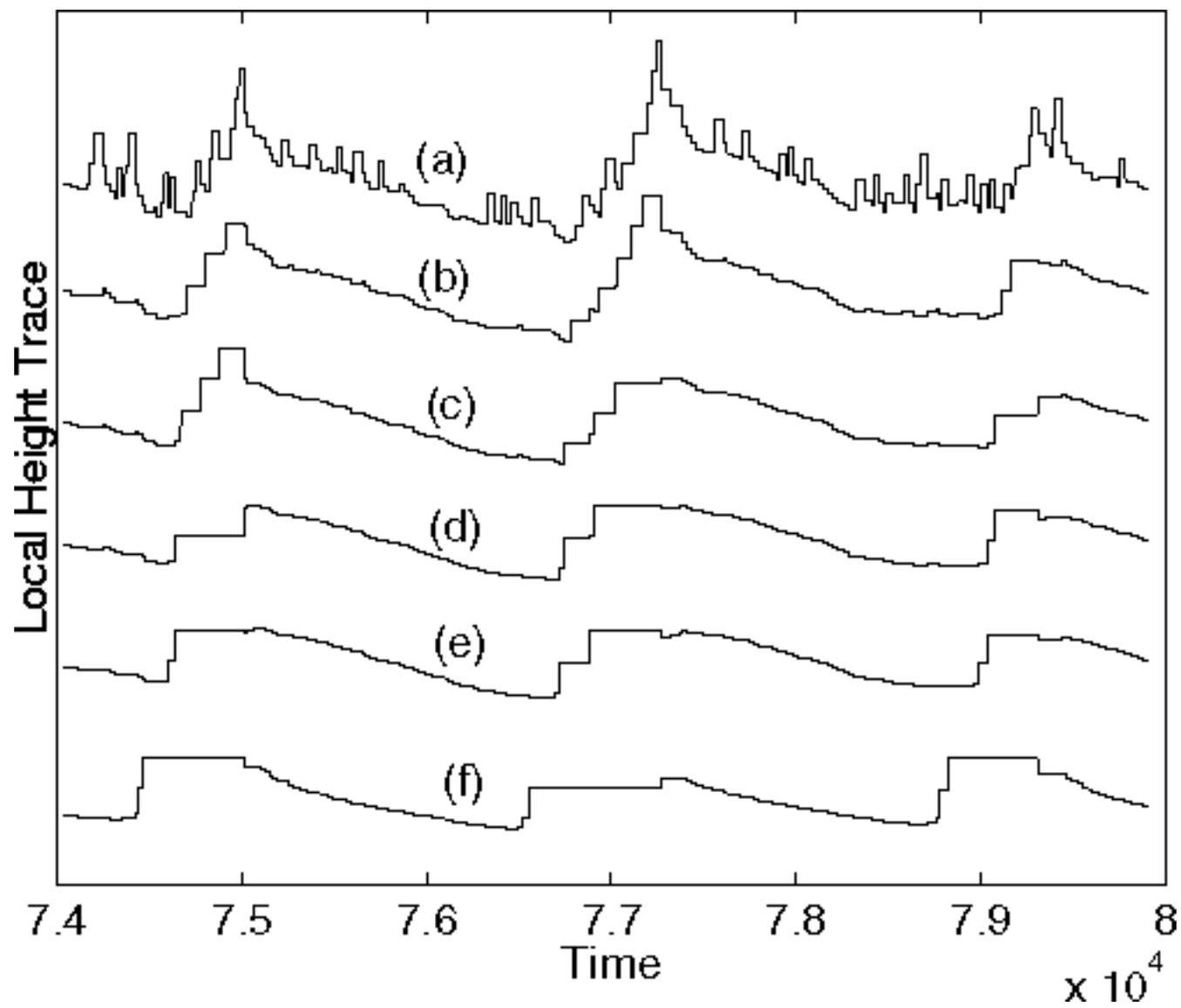
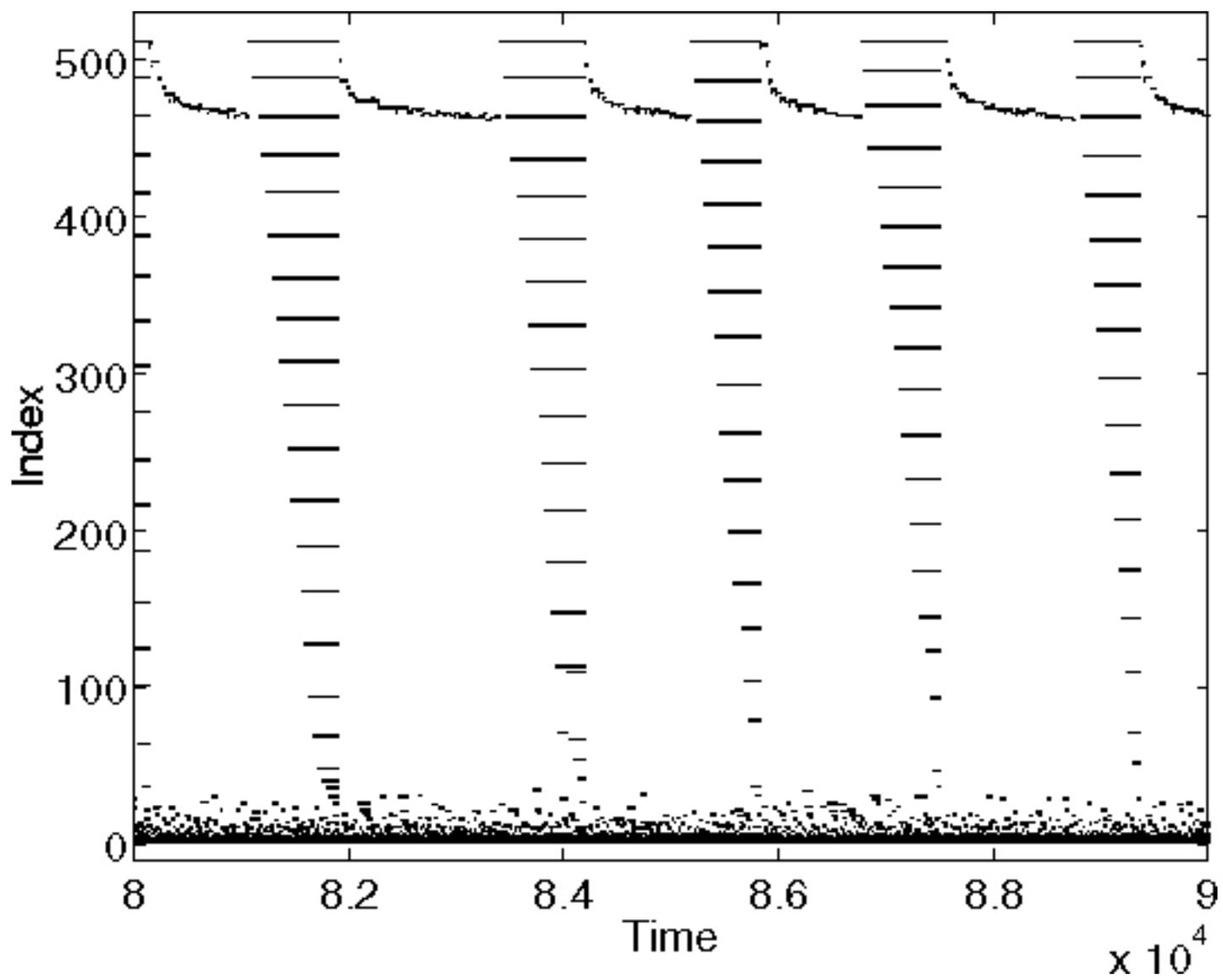
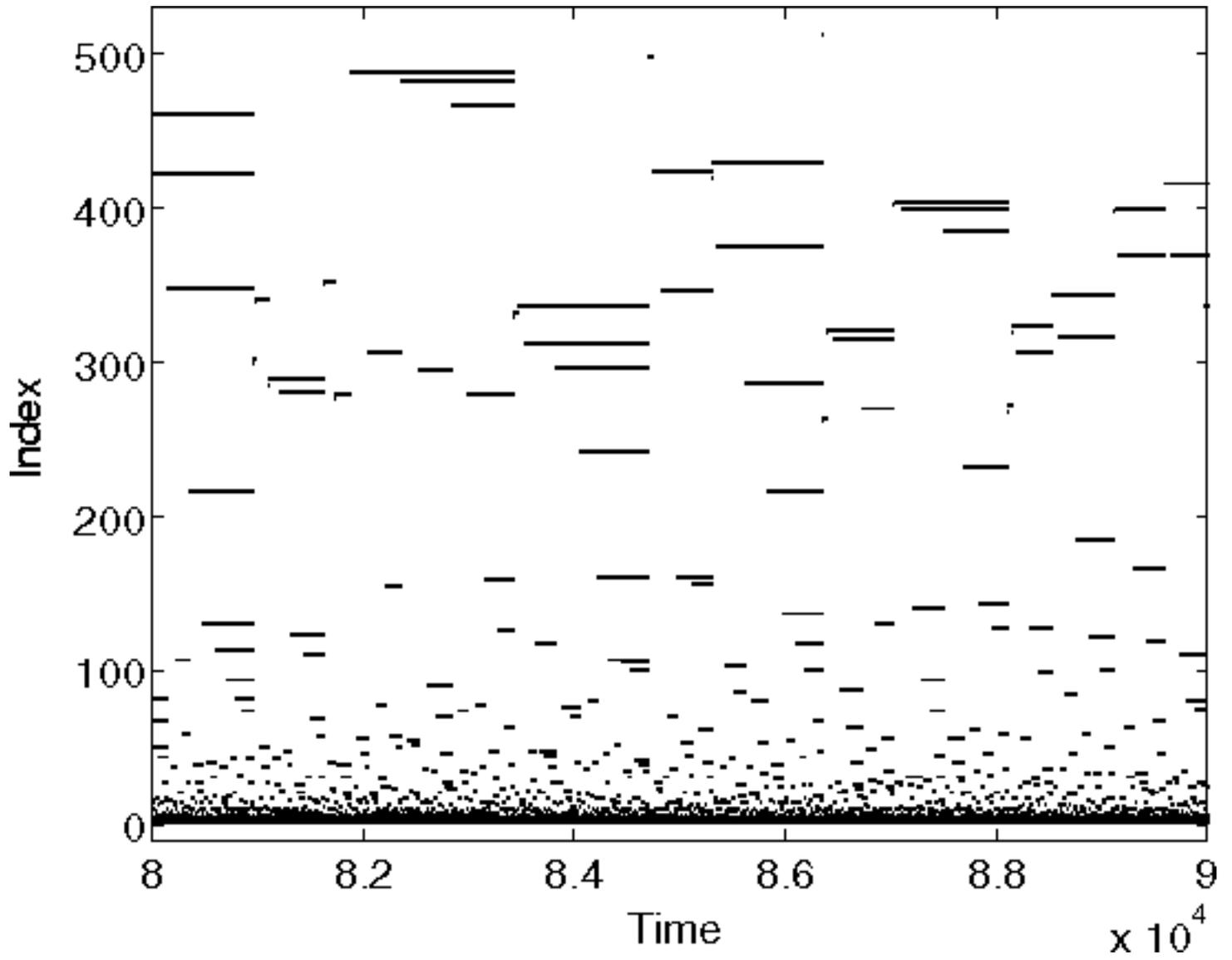
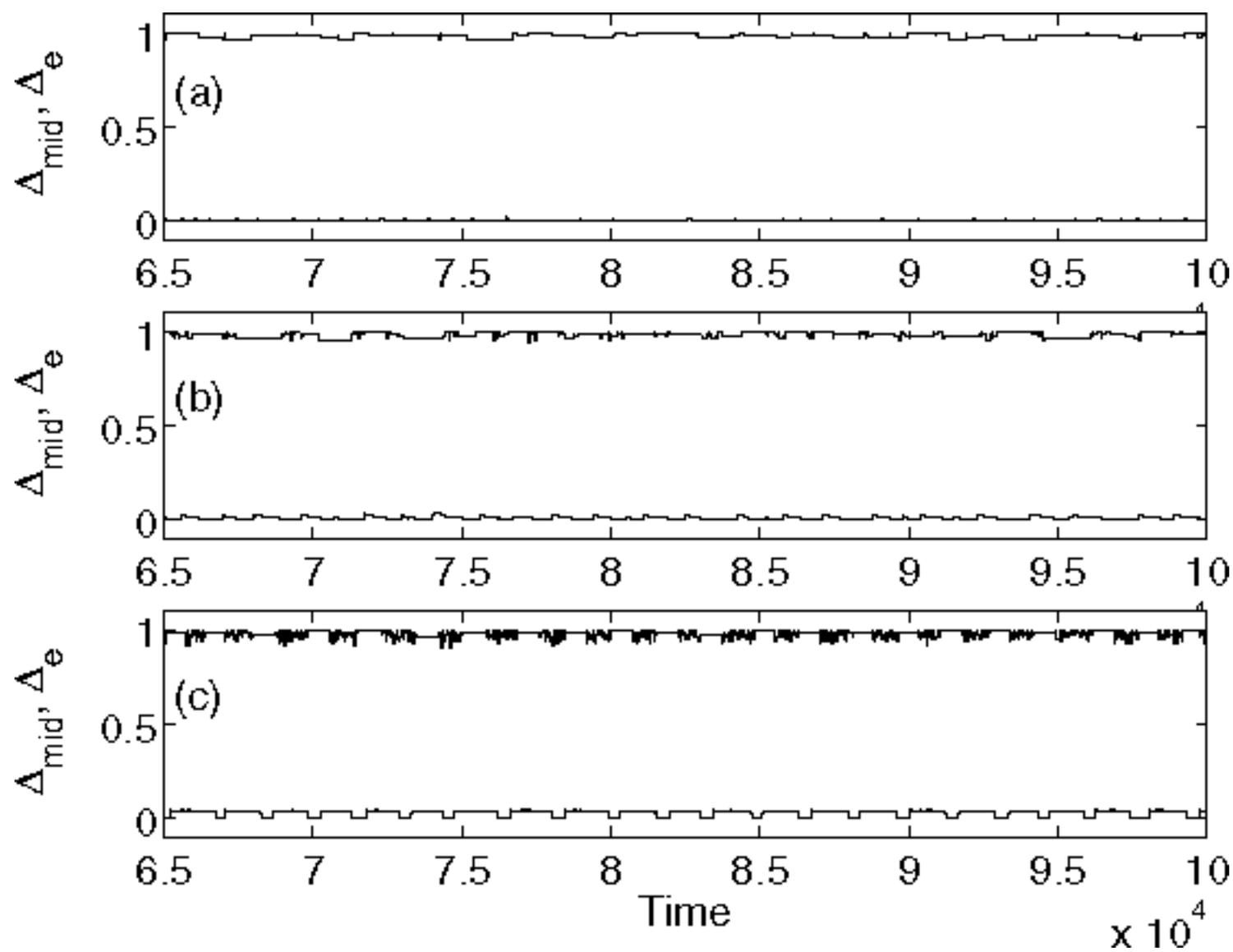
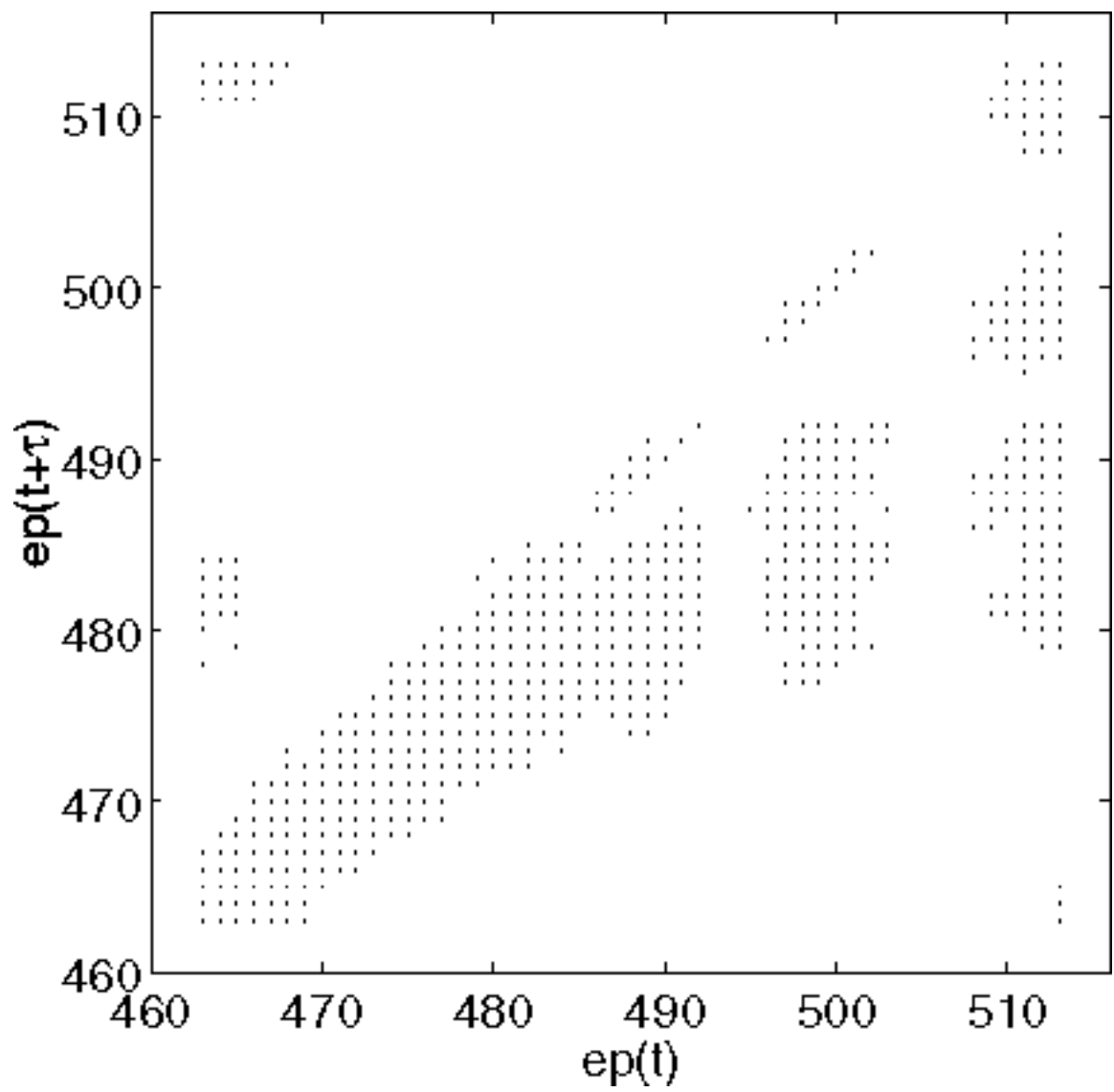
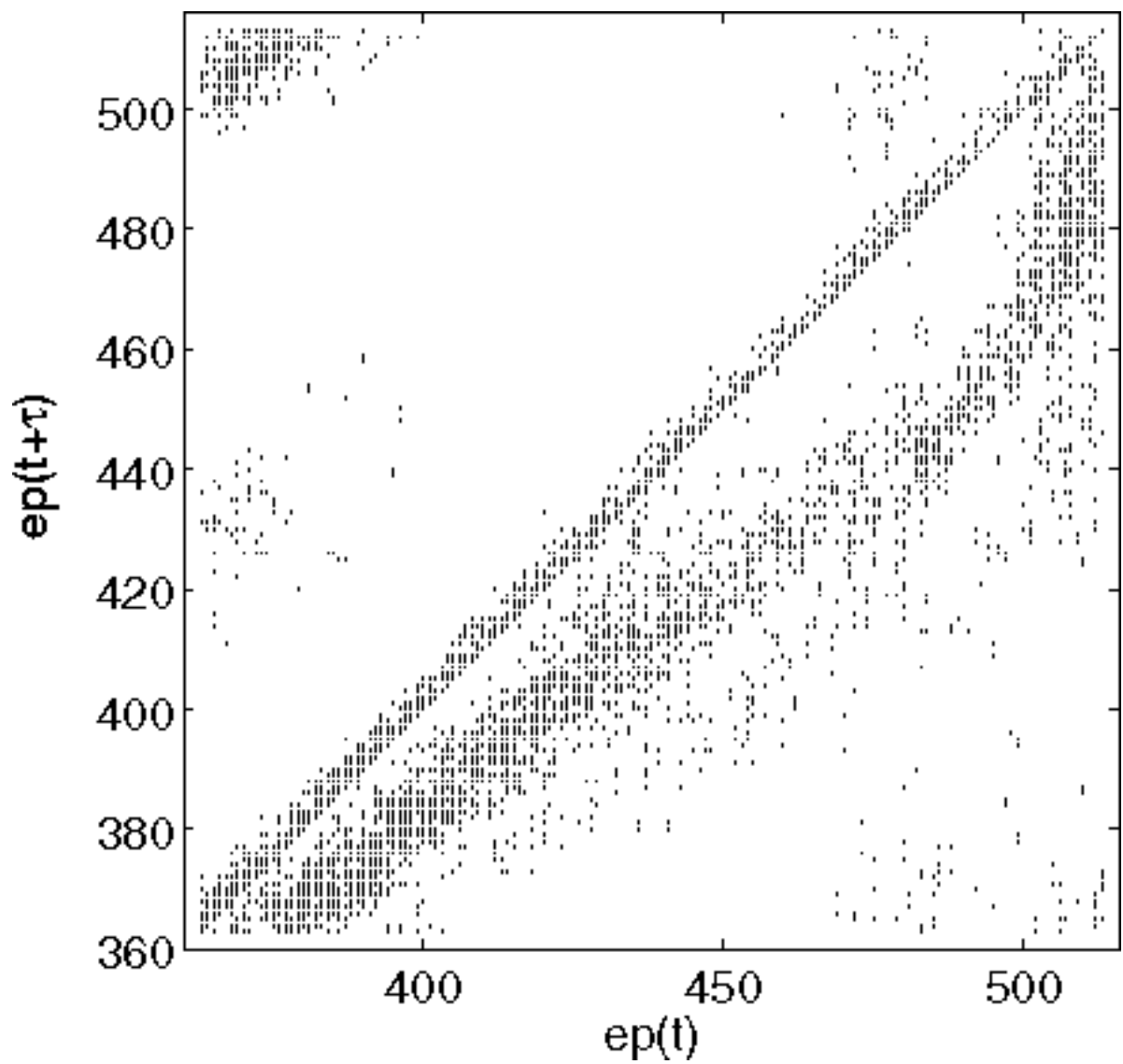
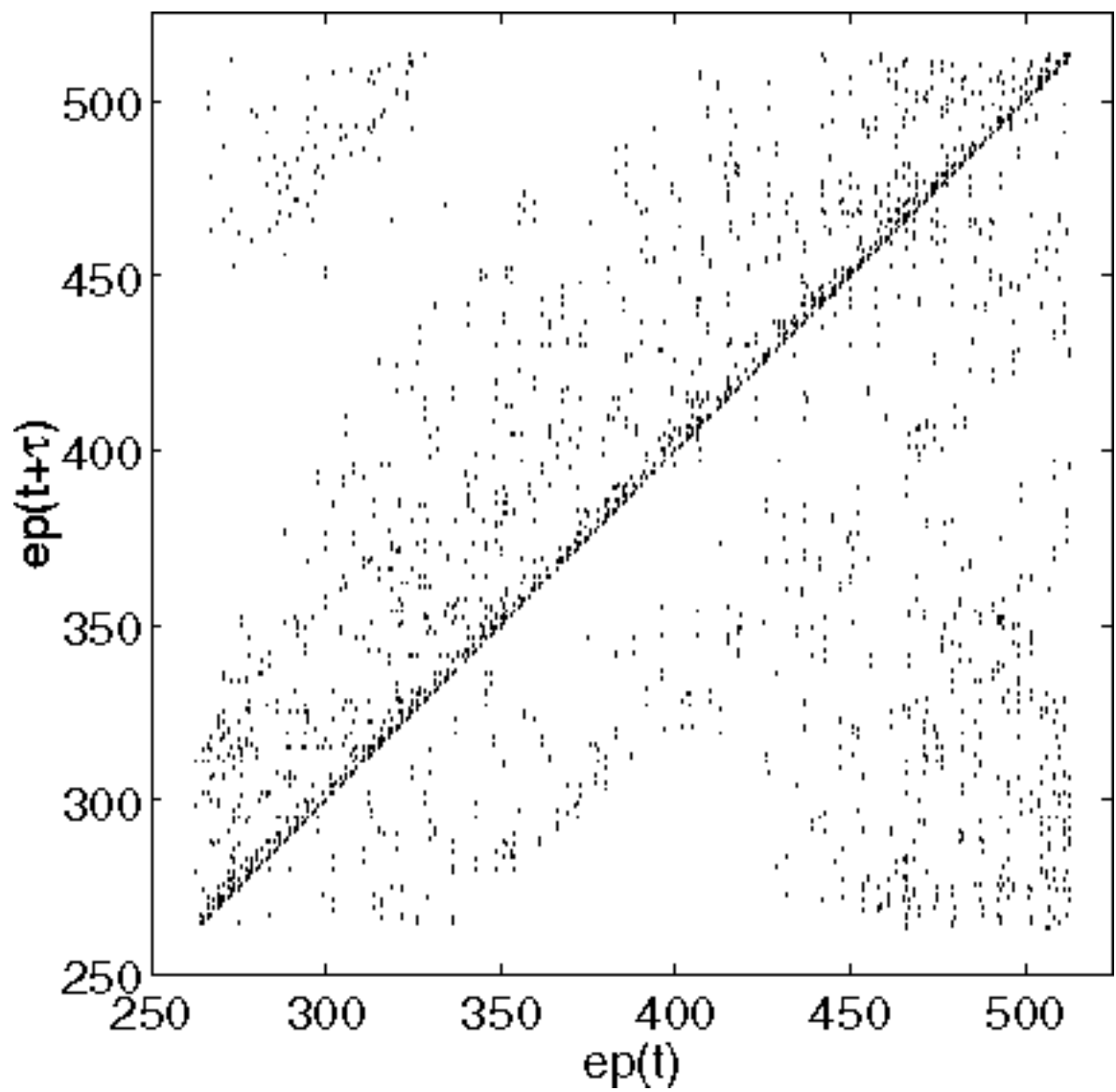
\end{document}